\documentclass[prl,twocolumn,aps,showpacs,floats,letterpaper,floatfix,groupedaddress]{revtex4}

\usepackage{graphicx,epsf, epsfig, amssymb}
\usepackage{bm}
\usepackage{longtable}

\def\be{\begin{equation}}
\def\ee{\end{equation}}
\def\beq{\begin{eqnarray}}
\def\eeq{\end{eqnarray}}

\begin{document}

\title{Black hole particle emission in higher-dimensional spacetimes}

\author{Vitor Cardoso} \email{vcardoso@phy.olemiss.edu}
\affiliation{Department of Physics and Astronomy, The University of
Mississippi, University, MS 38677-1848, USA \footnote{Also at Centro
de F\'{\i}sica Computacional, Universidade de Coimbra, P-3004-516
Coimbra, Portugal}}

\author{Marco Cavagli\`a} \email{cavaglia@phy.olemiss.edu}
\affiliation{Department of Physics and Astronomy, The University of
Mississippi, University, MS 38677-1848, USA}

\author{Leonardo Gualtieri} \email{gualtieri@roma1.infn.it}
\affiliation{
Centro Studi e Ricerche E. Fermi, Compendio Viminale, 00184 Rome, Italy and\\
Dipartimento di Fisica Universit\`a di Roma ``La Sapienza''/Sezione INFN Roma1, Piazzale Aldo Moro 2, 00185
Rome, Italy}


\begin{abstract}

In models with extra dimensions, a black hole evaporates both in the bulk and on the visible brane, where
standard model fields live. The exact emissivities of each particle species are needed to determine how the
black hole decay proceeds. We compute and discuss the absorption cross-sections, the relative emissivities and
the total power output of all known fields in the evaporation phase. Graviton emissivity is highly enhanced as
the spacetime dimensionality increases. Therefore, a black hole loses a significant fraction of its mass in the
bulk. This result has important consequences for the phenomenology of black holes in models with extra
dimensions and black hole detection in particle colliders.

\end{abstract}

\maketitle

A black hole is a grey body with temperature proportional to its surface gravity \cite{hawking}. The black hole
emission spectrum depends crucially on the structure and dimensionality of the embedding spacetime. In models
with large \cite{Arkani-Hamed:1998rs} or warped \cite{Randall:1999ee} extra dimensions, standard model fields,
except the graviton, are constrained to propagate on a four-dimensional submanifold of the higher-dimensional
spacetime. In this scenario, the relative emissivities of the fields (greybody factors) are essential to
determine if the black hole evaporates mainly on the four-dimensional brane or in the higher-dimensional bulk.
This is particularly important for models of low-energy scale gravity, where detection of subatomic black holes
in particle colliders and ultrahigh-energy cosmic ray observatories is only possible if a consistent fraction
of the initial black hole mass is channeled into brane fields \cite{Cavaglia:2002si}. Counting of brane vs.\
bulk degrees of freedom (d.o.f.) provides a naive argument in support of the dominance of brane over bulk
emission \cite{Emparan:2000rs}: Since all standard model fields carry a larger number of d.o.f.\ than the
graviton, brane emission must dominate over bulk emission. However, a large emissivity for bulk fields could
invalidate this conclusion \cite{Cavaglia:2003hg}. If the probability of emitting spin-2 quanta is much higher
than the probability of emitting lower spin quanta, the black hole may evaporate mainly in the bulk.

The relative emissivities per d.o.f.\ of a four-dimensional non-rotating black hole are 1, 0.37, 0.11 and 0.01
for spin-0, -1/2, -1 and -2, respectively \cite{page}. In that case, the graviton power loss is negligible
compared to the loss in other standard model channels. Since brane fields are constrained in four dimensions,
the relative greybody factors (relative emissivities) for these fields are expected to approximately retain the
above values in higher dimensions. The graviton emission is expected to be larger due to the increase in the
number of its helicity states. The helicity states of a massless particle in $D$-dimensions are given by the
representation of the little group $SO(D-2)$, i.e.\ the group of spatial rotations preserving the particle
direction of motion. For instance, a five-dimensional graviton has five helicity states, corresponding to the
$SO(3)$ group of orthogonal rotations to the direction of motion.

Therefore, a conclusive statement on brane vs.\ bulk emission rates requires
the knowledge of the greybody factors for all fields. The higher-dimensional
emission rates for spin-0, -1/2 and -1 fields are known \cite{kantireview}.
However, the exact emission rate for spin-2 fields has not been computed yet.
(See, however, Ref.\ \cite{CNS}). In this paper we fill this gap. We find that
the graviton emissivity is highly enhanced as the spacetime dimensionality
increases. Although this increase is not sufficient to lead to a domination of
bulk emission over brane emission, a consistent fraction of the
higher-dimensional black hole mass is lost in the bulk. This fact has important
consequences for the phenomenology of black hole events in low-scale gravity
and primordial black hole formation.

The $D$-dimensional non-rotating black hole is described by the
higher-dimensional Schwarzschild metric, also known as Tangherlini metric
\cite{TMP}
\be
ds^2= -fdt^2+ f^{-1}dr^2+r^2d\Omega_{D-2}^2\,,~~~~~f=1-\frac{1}{r^{D-3}}\,,
\label{metrictang}
\ee
where $d\Omega_{D-2}^{2}$ is the line element on the unit sphere $S^{D-2}$ and the black hole radius has been
set to $r_H=1$ without loss of generality. The formalism to handle gravitational perturbations of the
Tangherlini metric has been developed by Kodama and Ishibashi \cite{kodama} following previous work in four
dimensions by Regge and Wheeler \cite{regge} and Zerilli \cite{zerilli}. There are three kinds of gravitational
perturbations: scalar, vector and tensor. Scalar and vector perturbations have their counterparts in $D=4$. The
tensor type appears only in higher dimensions. The evolution equation for the perturbations is
\be
\frac{d^2\Psi}{dr_{*}^2}+(\omega^2-V)\Psi=0 \,,
\label{eveq}
\ee
where $r$ is a function of the tortoise coordinate $r_*$ defined by ${\partial
r}/{\partial r_*}=f$. For vector and tensor perturbations, the potential $V$ is
\be
V = f{\biggl [} \frac{l(l+D-3)}{r^2}+\frac{(D-2)(D-4)}{4r^2}+
\frac{(1-p^2)(D-2)^2}{4r^{D-1}} {\biggr ]} \,,
\label{potentialj}
\ee
where $l\ge 2,3,\dots$ is the angular quantum number and $p=2$ (0) for vector
(tensor) perturbations, respectively. For scalar perturbations, the potential
is
\be V= f\frac{Q(r)}{16r^2H(r)^2}\,,
\label{def-potential}
\ee
\begin{widetext}
\begin{eqnarray}
Q(r)&=&(D-2)^4(D-1)^2x^3+(D-2)(D-1)
\{4[2(D-2)^2-3(D-2)+4]m+(D-2)(D-4)(D-6)(D-1)\}x^2 \nonumber \\
&-&12(D-2)[(D-6)m+(D-2)(D-1)(D-4)]m x +16m^3+4D(D-2)m^2\,, \\
H(r)&=&m+{1\over 2}(D-2)(D-1)x\,,\quad m=l(l+D-3)-(D-2)\,,\qquad x=1/r^{D-3}\,.
\end{eqnarray}
\end{widetext}
Assuming a harmonic wave $e^{i\omega t}$ and ingoing waves near the horizon, we
have the following boundary condition:
\be \Psi(r)\rightarrow e^{- i\omega r_*} \sim (r-1)^{-
i\omega/(D-3)}\,,\quad r \rightarrow r_+\,.
\label{def-bcrmais}
\ee
For large $r_*$ we have both out- and in-going waves:
\be \Psi(r)\rightarrow T e^{-i\omega r_*}+ R e^{i\omega r_*}\,,\quad
r_*\rightarrow \infty\,.
\label{def-bc}
\ee
The absorption probability for the wave is $|{\cal
A}|^2=(|T|^2-|R|^2)/|T|^2$. The number of absorbed particles is
\be
\frac{dN_{abs}}{dt}=\sigma\Phi\,,
\label{defsigma}
\ee
where $\Phi$ is the flux and $\sigma$ is the absorption cross-section. A sum
over all final states and an average over initial states is understood. Solving
the equations for the gravitational perturbations with an expansion in
spherical harmonics, we find
\be
\sigma=C(D,\omega)\sum_l
\left[ N_{l\,S}\left|{\cal A}_{l\,S}\right|^2+
N_{l\,V}\left|{\cal A}_{l\,V}\right|^2
+N_{l\,T}\left|{\cal A}_{l\,T}\right|^2\right]\,,
\label{crossgrav}
\ee
where the subscripts $S$, $V$ and $T$ refer to scalar, vector and tensor
perturbations, respectively. The multiplicities $N_l$ are defined in Ref.\
\cite{multiplicities} and the normalization is
$C(D,\omega)=2(2\pi/\omega)^{D-2}[D(D-3)\Omega_{D-2}]^{-1}$.

The wave equation, Eq.\ (\ref{eveq}), can be solved in the low-energy regime
$\omega\ll 1$. (For details, see Ref.~\cite{inprep}.) This method uses a
matching procedure to find a solution valid throughout the whole spacetime for
any value of $p$. Low frequencies give a substantial contribution to the total
Hawking power emission, thus providing a good approximation of the exact
result. The absorption probability is
\begin{widetext}
\be
\left|{\cal A}\right|^2=1-|{\cal R}|^2=4\pi\left (
\frac{\omega}{2}\right )^{D+2l-2} \frac{\Gamma\left
(1+\frac{2l+p(D-2)}{2(D-3)}\right )^2\Gamma\left
(1+\frac{2l-p(D-2)}{2(D-3)}\right )^2}{\Gamma\left
(1+\frac{2l}{D-3}\right )^2\Gamma\left (l+\frac{(D-1)}{2}\right )^2}
\label{AbsCoeff}\,.
\ee
\end{widetext}
The low-energy absorption probability for spin-0 fields
\cite{kantimarchrussell} and spin-2 tensor perturbations is recovered by
setting $p=0$ in the above equation. The result for vector perturbations is
obtained by setting $p=2$. Gravitational scalar perturbations cannot be dealt
with analytically. However, numerical simulations give an effective $p_{\rm
grav\, scalar}\sim 2+0.674D^{-0.5445}$. The low-energy absorption cross section
can be obtained from Eqs.\ (\ref{AbsCoeff}). For instance, in
four-dimensions, where the tensor contribution disappears, the $l=2$ mode gives
the cross section $\sigma_{l=2}=4\pi\omega^4/45$. This result agrees with the
well-known result of Ref.\ \cite{page}.

The wave equation can also be solved in the high-energy limit. In that case, the absorption cross section is
expected to approximate the cross section for particle capture. This conjecture has been verified in a number
of papers for the scalar field. (See Ref.~\cite{kantireview} and references therein.) The proof for spin-1 and
spin-2 fields is sketched below. (For further details, see Ref.\ \cite{inprep}.) Since high frequencies can
easily penetrate the gravitational potential barrier, the absorption probabilities of all fields approximate 1
as $\omega \rightarrow \infty$. The cross section in the high-energy limit must include the contribution from
all $l \lesssim \omega$. Therefore, the largest contribution to the cross section is given by high-$l$ modes.
Since the high-$l$ limit of the wave equation is independent of the type of perturbation \cite{CDL03}, it
follows ${\cal A}_{l\,S}={\cal A}_{l\,V}={\cal A}_{l\,T}$. The universality of the high-energy absorption cross
section follows from Eq.\ (\ref{crossgrav}) and the properties of the multiplicities.

The total energy flux for the gravitational radiation is
\begin{widetext}
\be \frac{dE}{dt}=\frac{dE_S}{dt}+\frac{dE_V}{dt}+\frac{dE_T}{dt}=\sum_l \int \frac{d \omega}{2\pi} \frac{\omega}{e^{\omega/T_H}-1}\left (N_{l\,S} |{\cal
A}_{l\,S}^{s=2}|^2+N_{l\,V} |{\cal
A}_{l\,V}^{s=2}|^2+N_{l\,T} |{\cal
A}_{l\,T}^{s=2}|^2 \right )\,,
\label{totpower}
\ee
\end{widetext}
where we have separated the individual contributions of each harmonics. The
Hawking temperature is $T_H=(D-3)/(4\pi)$. Note that the number of helicities
is included in the multiplicity factors and the sum of scalar, vector and
tensor contributions.

The absorption probabilities ${\cal A}$ for all frequencies can be computed
numerically. Equation (\ref{potentialj}) is integrated from a point near the
horizon (typically $r-1 \sim 10^{-6}$), where the field behavior is given by
Eq.~(\ref{def-bcrmais}). The numerical result is compared to Eq.~(\ref{def-bc})
at large $r$. A better accuracy is achieved by considering the next-to-leading
order correction term (see Ref.\ \cite{bertihighd})
\be \Psi(r)\rightarrow T \left (1+\frac{\varepsilon}{r} \right )
e^{-i\omega r_*}+ R\left (1-\frac{\varepsilon}{r}\right ) e^{i\omega
r_*}\,,\quad r_*\rightarrow \infty\,.
\label{def-bc2}
\ee
This allows the determination of the coefficients $T\,,R$ and the absorption
probability ${\cal A}$. The results for the total integrated power, Eq.\
(\ref{totpower}), are summarized in Table \ref{tab:totalpower}. The values for
the lower-spin fields living on the brane are taken from Ref.\
\cite{kantireview}.
\begin{table*}[ht]
\caption{\label{tab:totalpower} Total power $P$ of Hawking radiation channeled
into different fields. The first three rows correspond to fields propagating on
the brane. The last row is the power radiated in bulk gravitons normalized to
the four-dimensional case.}
\begin{ruledtabular}
\begin{tabular}{ccccccccc}  \hline
$D$ &4&5&6&7&8&9&10&11\\ \hline
{\rm Scalars}&1 &8.94 &36&99.8 &222&429&749&1220\\ 
{\rm Fermions}&1 &14.2 &59.5&162 &352&664&1140&1830\\ 
{\rm Gauge Bosons}&1 &27.1 &144&441 &1020&2000&3530&5740\\ 
{\rm Gravitons}&1 &103 &1036& 5121 &$2\times 10^4$&$7.1\times
10^{4}$&$2.5\times 10^5$&$8\times 10^5$\\ 
\end{tabular}
\end{ruledtabular}
\vskip -2mm
\end{table*}
The graviton values are normalized to the four-dimensional case, where $P=1.52\times 10^{-5}$. This is in exact
agreement with Page's result. (See Table I of Ref.\ \cite{page}.) The results for lower-spin fields are
normalized to their four-dimensional values. It is worth discussing some features of the numerical results: (i)
The relative contribution of the higher partial waves increases with $D$. For instance, in four dimensions the
contribution of the $l=2$ mode is two orders of magnitude larger than the contribution of the $l=3$ mode. More
energy is channeled in $l=3$ mode than in $l=2$ mode for $D\ge 9$. (The largest tensor contribution in ten
dimensions comes from the $l=4$ mode.) Contributions from high $l$ are needed to obtain accurate results for
large $D$. For instance, in ten dimensions the first 10 modes must be considered for a meaningful result. The
values in Table \ref{tab:totalpower} have a 5\% accuracy. (ii) The total power radiated in gravitons increases
more rapidly than the power radiated in lower-spin fields as $D$ increases. This is due to the increase in the
multiplicity of the tensor perturbations, which is larger than the scalar multiplicity by a factor $D^2$ at
high $D$. Therefore, the main contribution to the total power comes from the tensor (and vector) modes. For
instance, in ten dimensions the tensor mode contributes roughly half of the total power output.

Table II gives the fraction of radiated power per d.o.f.\ normalized to the
scalar field, where the graviton value includes all the helicity states. In
four dimensions, the power loss in gravitons is negligible compared to the
power loss in lower-spin fields. The graviton channel is only about 5\% of the
scalar channel. This conclusion is reversed in higher dimensions. For instance,
the graviton loss is about 35 times higher than the scalar loss in $D=11$.
Although the graviton emission is highly enhanced, the large number of brane
d.o.f.\ (more than 100 for the standard model) assures that the brane
channel dominates on the bulk channel. However, power loss in the bulk is
significant and cannot be neglected at high $D$; about 1/4 of the initial black
hole mass is lost in the 11-dimensional bulk.
\begin{table}[h]
\caption{\label{tab:totalpower3} Fraction of radiated power per d.o.f.\
normalized to the scalar field. The graviton d.o.f.\ (number of
helicity states) are included in the results.}
\begin{ruledtabular}
\begin{tabular}{ccccccccc}  \hline
$D$ &4&5&6&7&8&9&10&11\\ \hline
{\rm Scalars}        &1      &1    &1    &1   &1   &1  &1&1\\ 
{\rm Fermions}       &0.55   &0.87   &0.91   &0.89    &0.87  &0.85  &0.84&0.82\\ 
{\rm Gauge \, Bosons}&0.23    &0.69   &0.91   &1.0  &1.04  &1.06  &1.06&1.07\\ 
{\rm Gravitons }      &0.053   &0.61    &1.5    &2.7   &4.8   &8.8   &17.7 &34.7\\ 
\end{tabular}
\end{ruledtabular}
\vskip -2mm
\end{table}
The particle emission rates per d.o.f.\ are shown in Table III. The relative emission rates of different fields
can be obtained by summing on the brane d.o.f.\ For instance, the relative emission rates of standard model
charged leptons (12 d.o.f.) and the 11-dimensional bulk graviton are roughly 1:1. This ratio becomes $\sim$
40:1 in five dimensions. We find again that the bulk energy loss is significant in higher-dimensional
spacetimes.
\begin{table}[h]
\caption{\label{tab:emission rates} Fraction of emission rates per d.o.f.\
normalized to the scalar field. The graviton result includes all the helicity
states and counts as one d.o.f.}
\begin{ruledtabular}
\begin{tabular}{ccccccccc}  \hline
$D$ &4&5&6&7&8&9&10&11\\ \hline
{\rm Scalars}        &1      &1    &1    &1   &1   &1  &1&1\\ 
{\rm Fermions}       &0.37   &0.7   &0.77   &0.78    &0.76  &0.74  &0.73&0.71\\ 
{\rm Gauge \, Bosons}&0.11    &0.45   &0.69   &0.83  &0.91  &0.96  &0.99&1.01\\ 
{\rm Gravitons }    &0.02    &0.2   &0.6  &0.91   &1.9   &2.5 &5.1  &7.6\\ 
\end{tabular}
\end{ruledtabular}
\vskip -2mm
\end{table}
This long overdue computation is of paramount importance for the phenomenology
of any higher-dimensional gravitational models. Our main result is that the
power loss in the graviton channel is highly enhanced in higher-dimensional
spacetimes. This has important consequences for the detection of microscopic
black hole formation in particle colliders and ultrahigh-energy cosmic ray
observatories, where a larger bulk emission implies larger missing energies and
lower multiplicity in the visible channels. Despite the increase in graviton
emissivity, a non-rotating black hole in the Schwarzschild phase will emit
mostly on the brane due to the higher number of brane d.o.f. However, black
hole energy loss in the bulk cannot be neglected in presence of extra
dimensions.

The effects of graviton emission may be even more dramatic when one includes
rotation. Consider a rotating four-dimensional black hole \cite{page2}.
Graviton emission, which is supressed for small rotations, rapidly increases
with the angular momentum $J$. (In four dimensions $J$ ranges from $0$ to
$M^2$). As $J$ grows from $0$ to $0.7M^2$ graviton emissivity grows 3 orders of
magnitude (see Table I in Ref.\ \cite{page2}) while the emissivities of fermion
and gauge bosons grow less than one order of magnitude. A similar behavior is
expected in higher dimensions, where for $D>5$ there is no upper bound on $J$,
implying that graviton emission dominates the evaporation process. Since known
results for rotating black holes in $D$-dimensions do not include gravitons
\cite{HawkingHigherDimRot}, this remains an open question.

\noindent{\bf Acknowledgements.} We are very grateful to De-Chang Dai for pointing out some typos in previous
versions. We would like to thank Akihiro Ishibashi for useful correspondence. VC acknowledges financial support
from FCT through PRAXIS XXI program, and from Funda\c c\~ao Calouste Gulbenkian.

\end{document}